# Oriented Polymers: A Transfer Matrix Calculation

W.M.Koo[*]

Center for Theoretical Physics

Seoul National University

Seoul 151-742, Korea[†]

9 December 1994

## Abstract

Based on transfer matrix techniques and finite size scaling, we study the oriented polymer (self-avoiding walk) with nearest neighbor interaction. In the repulsive regime, various critical exponents are computed and compared with exact values predicted recently. The polymer is also found to undergo a spiral transition for sufficiently strong attractive interaction. The fractal dimension of the polymer is computed in the repulsive, attractive regimes and at the spiral transition point. The later is found to be different from that at the collapse transition of ordinary self-avoiding walk.

[*]Work supported in part by a grant from the Korea Science and Engineering Foundation through the SRC program of SNU-CTP, and the Ministry of Education.

[†]SNUTP-94-127

0

# 1 Introduction

A polymer chain describes a self avoiding walk (SAW). In the high temperature phase, one generally ignores the Van Der Waals attraction between monomers of the polymer chain and study the polymer problem as the $n \to 0$ limit of the $O(n)$ model[1]. In 2d, this has been the basis for calculation of many critical exponents that have important physical meaning[1, 2, 3, 5], and also starting point for many interesting extensions of the polymer problem such as polymer in the low temperature phase where monomers' attractions are taken into account[4, 5, 6, 7]. However, much less attention has been paid to the study of polymer chain with orientation, ie. polymer with an 'arrow' that runs along the chain[8, 9]. Physically, one can imagine such orientation as arising from dipole moments of the monomers that make up the chain. Clearly, if we only consider the interactions between monomers to be the excluded volume ones, then the orientation does not play any role and there is no difference between the oriented and ordinary polymers. Indeed, an oriented polymer chain can be considered as the $n \to 0$ limit of the *complex* $O(n)$ model whose partition function is given by,

$$\sum_{\{\vec{S}\}} \prod_{<ij>} (1 + x\vec{S}_i \cdot \vec{S}_j^* + x\vec{S}_i^* \cdot \vec{S}_j) \qquad (1.1)$$

where $\vec{S}_i$ is a complex $n$-component vector at site $i$ with $*$ denotes complex conjugation, $<>$ denotes nearest neighbor lattice sites, and $x$ is the fugacity of the monomer. The complex $O(n)$ model is the same as the $O(2n)$ model, hence the oriented SAW is in the same universality class as the ordinary SAW.

An interesting extension of this model consists in introducing interaction between the monomers that depends on their relative orientation[9]. More precisely, consider two monomers that are in close encounter, (for a square lattice, for example, this can be defined by neighboring lattice edges being occupied by the monomers) their relative orientation can either be parallel or anti-parallel. We assign different Boltzmann weights 1 or $e^{\lambda_o}$ to these two types of close encounters (see fig. 1). Note that $\lambda_o$ here differs from that in [9] by a sign.

Such an orientation dependent interaction has been identified in the context of the complex $O(n)$ model as a current-current interaction[9]. The complex $O(n)$ model possesses a $U(1)$ symmetry associated with $\vec{S} \to e^{i\alpha}\vec{S}$, the corresponding current $J_\mu = \frac{i}{2}(\vec{S}^* \cdot \partial_\mu \vec{S} - \vec{S} \cdot \partial_\mu \vec{S}^*)$ can be identified as the tangent vector along the oriented chain[8]. Perturbing the complex $O(n)$ model by a current-current interaction, $\lambda \int J\bar{J} dz d\bar{z}$, corresponds to introducing an orientation dependent interaction for the monomers of the polymer chain.

This model includes the ordinary SAW as a special case where $\lambda_o = 0$. For $\lambda_o \ll 0$, parallel close encounters are suppressed. While for $\lambda_o \gg 0$ parallel close encounters are preferred, the polymer has the tendency to coil up to form a spiral. We shall denote these two phases respectively as the *repulsive* and *attractive* regimes. For some intermediate $\lambda_o > 0$ we expect a transition to take place. This transition, denoted as the *spiral transition*, should be different from the usual collapse transition (or the $\theta(\theta')$ point) of the ordinary SAW[5, 6, 7] as one can readily see that in this case it is the self-trapping configurations that are responsible for the collapse. We therefore expect a new multi-critical point from this spiral transition.

On the other hand, in the repulsive regime, regarding the current-current interaction as a perturbation from the ordinary SAW, which belongs to this regime, and using the fact that the perturbation is truly marginal[10], critical exponents can be determined exactly as functions of $\lambda$ (the renormalized $\lambda_o$)[9]. For example, the operator which corresponds to the source of a $L$-leg oriented SAW with $U(1)$ charge $L$, (ie. $L$ arrows originate from the source) scales with exponent

$$x_L(\lambda) = x_L(0) - 2\pi\lambda L^2 \tag{1.2}$$

where

$$x_L(0) = \frac{9L^2 - 4}{48} \tag{1.3}$$

is the critical exponent for the ordinary SAW. Also in this case, the fractal dimension of the SAW is still given by that of the ordinary SAW, $D_F = \frac{4}{3}$, since the energy operator $\vec{S}_i \cdot \vec{S}_j^* + \vec{S}_i^* \cdot \vec{S}_j$ is associated with self avoiding ring, which can not have parallel close encounter.

In this paper, we present a numerical study of this interacting oriented polymer problem based on transfer matrix and finite size scaling techniques. We mainly consider oriented SAWs on a square lattice with periodic boundary condition. For the repulsive regime, we verify the various critical exponents $x_L(\lambda)$ conjectured, and demonstrate the presence of a spiral transition as one increases $\lambda_o$. On the spiral transition point, exponents $\nu_x$ and $\gamma$ are determined numerically.

The basic quantities that we are interested are the connectivity constant and the critical exponents. They can be determined from the spin-spin correlation function. The spin-spin correlation function in the high temperature expansion can be expressed as a sum over all configurations of oriented SAW with source and sink at the locations 0 and $\vec{R}$ of the two spins respectively, such that each configuration is weighted by the length and the number of parallel close encounters as given below:

$$< \vec{S}(0) \cdot \vec{S}^*(\vec{R}) > = \sum_{\mathcal{C}} x^{l(\mathcal{C})} e^{\lambda_o m(\mathcal{C})} . \qquad (1.4)$$

Here $l(\mathcal{C})$ and $m(\mathcal{C})$ are respectively the length ( number of steps ) and the number of parallel close encounters of the oriented SAW $\mathcal{C}$. Note that in this model we take the weight of anti-parallel close encounter to be 1 as this choice ensures that the critical fugacity is independent of $\lambda_o$ in the repulsive regime[9], to which we shall mainly restrict our attention.

The expansion can be rewritten as

$$< \vec{S}(0) \cdot \vec{S}^*(\vec{R}) > = \sum_l Z_l(0, \vec{R}) x^l \qquad (1.5)$$

where

$$Z_l(0, \vec{R}) = \sum_m \omega_{l,m}(0, \vec{R}) e^{\lambda_o m} \qquad (1.6)$$

is defined as the partition function of a polymer chain. Hence the correlation function is a generating function of the polymer partition function. In the vicinity of the critical point, $x \to x^{c-}$,

$$< \vec{S}(0) \cdot \vec{S}^*(\vec{R}) > \sim \frac{1}{|\vec{R}|^{2x_1}} e^{-|\vec{R}|/\xi} \quad |\vec{R}| \to \infty \tag{1.7}$$

where $x_1$ is the exponent given in eqn.( 1.2) and the correlation length $\xi$ diverges as $|x - x^c|^{-\nu}$. The constants $x^c$, $x_1$ and $\nu$ have geometrical meanings as can be seen by inverting the above generating function. One readily shows that the SAW partition function has the asymptotic form

$$Z_l(0, \vec{R}) \sim (x^c)^{-l} l^{\gamma(\lambda)-1-2\nu} F(|\vec{R}|/l^\nu) \quad l, |\vec{R}| \to \infty \tag{1.8}$$

where

$$\gamma(\lambda) = \nu(2 - 2x_1(\lambda)) \tag{1.9}$$

and $F$ is some scaling function. Hence, $x^{c-1}$ is the connectivity constant that depends on the type of lattice, which by our choice of the interaction energy, does not depend on $\lambda$ ($\lambda_o$). For a square lattice, the value has been determined with high precision to be $0.3790528(25)$[11]. The thermal exponent which is also independent of $\lambda$ is equal to the inverse of the fractal dimension of the model since the average radius of gyration

$$< |\vec{R}| > = \sum_{\vec{R}} |\vec{R}| Z_l(0, \vec{R}) / \sum_{\vec{R}} Z_l(0, \vec{R}) \tag{1.10}$$

behaves asymptotically as

$$< |\vec{R}| > \sim l^{1/\nu} \quad l \to \infty. \tag{1.11}$$

So geometrically, $(x^c)^{-1}$ and $\gamma$ determine the number configurations of a polymer with given length, while $\nu$ determines the average size.

Besides the spin-spin correlation function, one can also consider correlation function of an operator $\phi_L$ with $U(1)$ charge $L$ that corresponds to insertion of the source of a $L$-leg oriented SAW. In this case, the correlation function generates the 'watermelon' diagram consisting of $L$ mutually and self avoiding oriented walks tied together at their extremities.

## 2 Polymer on a strip

Consider a polymer on a strip of finite width $n$. The generating function of $Z_l(0, \vec{R})$, and hence the constants $x^c$, $\gamma$, and $\nu$, can be obtained from the transfer matrix. The method employed is similar to that used in the study of ordinary polymer[11, 3, 5]. A state is still characterized by monomers' configurations on a column of the strip as well as the connectivity to the left of these monomers. But in our case, monomers and their connectivity carry orientations. As a result, besides the usual non-crossing constraint, connectivity has to satisfy additional constraint which may be regarded as the non-crossing constraint for connectivity to the right of the monomers. As an example, the column of monomers and their connectivity shown in fig. 2 can not represents section of a *single* SAW.

To keep track of the monomers' interactions, one has to consider monomers' configurations on a column of horizontal and vertical edges, this together with the fact that connectivity carries orientation vastly increases the number of possible states. In most cases we manage to diagonalize transfer matrix of dimension of the order of $7000 \times 7000$, which for the one polymer sector is equivalent to a strip of size 7 with translational symmetry taken in account. We list some of their dimensions in table 1 where $S_{L,n}$ denotes dimension of the translational invariant sector with $L$ polymers and $n$ is the width of the strip.

For large $|\vec{R}|$, the generating function eqn.( 1.4) is given by the largest eigenvalue $\Lambda$ of the transfer matrix of the one polymer sector as $\Lambda^{|\vec{R}|}$ and this defines a correlation length $\xi_n(x, \lambda_o)$ by

$$\Lambda^{|\vec{R}|} = e^{-|\vec{R}|/\xi_n}. \qquad (2.1)$$

The critical point is defined as the smallest $x_n^c$ where the correlation length diverges (ie. $\Lambda(x_n^c, \lambda_o) = 1$). Keeping only the dominating term in eqn.( 1.8), the free energy is simply given by

$$f_n(\lambda_o) = -\log x_n^c(\lambda_o) \qquad (2.2)$$

from which one can determine all thermodynamics quantities. In particular, the susceptibility with respect to $\lambda$ can be obtained as

$$c_n(\lambda_o) = -\frac{d^2}{d\lambda_o^2} \log x_n^c \qquad (2.3)$$

We present in fig. 3 a plot of this susceptibility against $\lambda_o$ for various strip size $n$ ranging from 3 to 7. This susceptibility exhibits a peak near $\lambda_o \sim 1$, signaling a possible phase transition.

Using the partition function eqn.( 1.8), the average radius of gyration of a polymer chain of length $l$ can be expressed as

$$<|\vec{R}_l|> = l \left( \frac{\partial}{\partial \log x} \log \Lambda(x_n^c, \lambda_o) \right)^{-1} . \qquad (2.4)$$

Since there is only one polymer in the area $n <|\vec{R}_l|>$, the average density of the chain is thus given by

$$\rho_n(\lambda_o) = \frac{1}{n} \frac{\partial}{\partial \log x} \log \Lambda(x_n^c, \lambda_o) . \qquad (2.5)$$

This density is plotted in fig. 4, it increases with $\lambda_o$, where $\lambda_o \sim 1$ divides the low and high density regimes. Since polymer in the repulsive (attractive) regime has low (high) density, $\lambda_o \sim 1$ is roughly the spiral transition point. This change of density can also be seen from the coefficient of expansion defined as

$$t_n(\lambda_o) = -\frac{1}{\rho_n} \frac{d\rho_n}{d\lambda_o}. \qquad (2.6)$$

In fig. 5, We again notice a peak around $\lambda_o \sim 1$.

From these figures, we see strong evidence of a singularity near $\lambda_o \sim 1$ building up as $n \to \infty$. In table 2 locations of successive maximums of $c_n$ and $t_n$ are listed and these values converge to $1.17\pm0.2$ and $1.12\pm0.2$ respectively. It is very likely that this is the location of the spiral transition point. We shall denote this point as $\lambda_o^*$.

So far we have concentrated on the critical line, but as in the case of ordinary polymer, one can give physical meaning to the entire $(x, \lambda_o)$ plane by considering the polymer as a chain with flexible length controlled by the fugacity $x$, the analysis is essentially similar to that of the ordinary SAW[3, 5]. In summary, every point in the $(x, \lambda_o)$ plane is accessible to the polymer chain, the critical line $x = x^c(\lambda_o)$ now corresponds to the isobar where the pressure vanishes. Polymer in the region below (above) this line is subjected to negative (positive) pressure and is swollen (dense). It can also be shown that crossing this critical line in the attractive regime one encounters a second order phase transition, while crossing it in the attractive regime one encounters a first order phase transition characterized by a jump in the density defined as

$$\rho_n(x, \lambda_o) = \frac{1}{n}\frac{\partial}{\partial \log x} \log \Lambda(x, \lambda_o) \qquad (2.7)$$

in this grand canonical picture.

## 3  Thermal exponent $\nu$

The correlation function $\xi(x, \lambda_o)$ diverges at the critical line $x = x^c$. For given $\lambda_o$, the divergence is characterized by the exponent $\nu$. In the repulsive regime, the thermal exponent $\nu$ is expected to be equal to that of the ordinary SAW ie. 3/4. While in the attractive regime, because the polymer is dense, it is expected to be equal to 1/2. Note that this value is also expected from the first order transition nature of the critical line in this regime[12]. The spiral transition point, which divides these two regimes, has $\nu \equiv \nu_x$ assumes some intermediate value. Near $x^c$, the correlation length $\xi_n$ defined in eqn.( 2.1) has the following finite size scaling form

$$\xi_n(x, \lambda_o) \sim n F_\xi\left((x - x^c)n^{1/\nu}\right) \qquad n \gg 1, x - x^c \ll 1. \qquad (3.1)$$

From eqn.( 2.5) and $\log \Lambda = -1/\xi_n$, it is clear that

$$\nu_n \equiv \left( \frac{\log(\rho_{n+1}(x^c_{n+1}, \lambda)/\rho_n(x^c_n, \lambda))}{\log(n+1/n)} + 2 \right)^{-1} \qquad (3.2)$$

converges to $\nu$ as $n \to \infty$.

In fig. 6 we plot successive values of $\nu_n$ against $\lambda_o$. The various lines for different strip sizes intersect almost at the same point at $\lambda_o \sim 1$ where spiral transition takes place. Below this point, the thermal exponent converges to 3/4 rather slowly, while above this point, $\nu$ decreases to 1/2 as expected from earlier prediction.

An alternative method to determine this exponent uses the phenomenological renormalization relation

$$\frac{1}{n}\xi_n(\tilde{x}^c_n, \lambda_o) = \frac{1}{m}\xi_m(\tilde{x}^c_n, \lambda_o) \qquad n, m \gg 1 \qquad (3.3)$$

which is again based on the finite size scaling assumption eqn.( 3.1). Note that the above relation is also used in the first order transition line where $\lambda_o > \lambda^*_o$, this is justified by the fact that the correlation length diverges at the critical point. The extrapolated critical value $\tilde{x}^c$ for large $n$ obtained from this relation can be checked to be in agreement with that obtained from the jump of the density $\rho_n(x, \lambda_o)$.

In the numerical calculation we take $m = n+1$ in eqn.( 3.3) where convergence is fastest to determine successive critical line $x = \tilde{x}^c_n$. The thermal exponent can be determined from the derivative of the correlation length by extrapolating $\nu_n$ defined now as

$$\nu_n \equiv \left( \log \left( \frac{\partial \xi_{n+1}(\tilde{x}^c_n, \lambda_o)/\partial x}{\partial \xi_n(\tilde{x}^c_n, \lambda_o)/\partial x} \right) / \log \left( \frac{n+1}{n} \right) - 1 \right)^{-1}. \qquad (3.4)$$

The result shown in fig. 7 again shares the same features of the previous figure: Thermal exponents $\nu$ extrapolated in the attractive and repulsive regimes are compatible with that obtained before but in this case the convergence is slightly faster. Furthermore, various lines cross also at $\lambda_o \sim 1$.

Successive $x_n^c$ obtained from eqn.( 3.3) are shown in fig. 8. Indeed, the extrapolated $\tilde{x}_n^c$ for large $n$ is not sensitive to $\lambda$ in the section of the repulsive regime where $\lambda_o < \lambda_o^*$ as expected from the choice of the relative weight for the monomers' interactions. So indeed the critical fugacity is equal to that of the ordinary SAW

$$x^c(\lambda_o) = 0.3790528(25) \ . \qquad (3.5)$$

Another interesting feature of these thermal exponents is the crossings of the various lines, which converge to the spiral transition point as $n \to \infty$. The corresponding thermal exponents and $\lambda_o$ values for various crossing of the lines are listed in table 3. Due to the small number of data and system sizes we have, it is difficult to obtain any reliable extrapolated values. Nonetheless, from both sets of data, it is likely that

$$\begin{aligned} \lambda_o^* &\simeq 1.1 \pm 0.2 \\ \nu_x &\simeq 0.63 \pm 0.05 \end{aligned} \qquad (3.6)$$

It should be remarked that since the free energy (eqn.( 2.2)) is always greater than the free energy associated with a tightly wound spiral (which is given by $\lambda_o$), and that $x^c$ in the entire repulsive regime up to the spiral transition point is given by eqn.( 3.5) independent of $\lambda_o$. We have

$$\log(x^c)^{-1} > \lambda_o \qquad \text{for } \lambda_o \leq \lambda_o^* \ . \qquad (3.7)$$

This implies the following bound

$$\lambda_o^* \leq 0.970080(7) \ . \qquad (3.8)$$

The values for $\lambda_o^*$ deduced from tables 2, 3 are very close to this bound. The fact that they exceed this bound can be attributed to the small system sizes we can handle and hence the unreliability of the extraploted values. Note that the estimated $\lambda_o^*$ is very close to the value of the bound ie. $\log(x^c)^{-1}$. It is tempting to conjecture that indeed the $\lambda_o^*$ is given by the bound ie. the value obtained from the critical fugacity of ordinary SAW. If this is indeed the case then this implies that there is no $O(l)$ excitation from the tightly wound spiral, which suggests that the spiral transition is first order and quite different from the $\theta$ point for the collapse of ordinary SAW. We hope to pursue this issue in the future.

# 4   Conformal Weights

In the repulsive regime the exponent $x_L(\lambda)$ that characterizes the scaling behaviors of the $L$-leg oriented SAW can be related to that of the ordinary SAW. The latter has been identified as conformal weight of a twisted $N = 2$ supersymmetry theory with $k = 1$[14]. Thus, the oriented SAW in this regime is described by a marginal perturbation of the twisted $N = 2$ theory in the continuum limit[15]. This exponent can likewise be computed using the transfer matrix technique. As in eqn.( 1.4), we consider the correlation function of the $L$-leg operator $\phi_L$ on a finite strip of width $n$ and define a correlation length $\xi_{L,n}$ from the largest eigenvalue of the corresponding transfer matrix as

$$\Lambda_{L,n} = e^{-1/\xi_{L,n}} . \tag{4.1}$$

From conformal field invariance argument[17], $x_{L,n}(\lambda_o)$ defined as

$$x_{L,n}(\lambda_o) = \frac{n}{\pi \xi_{L,n}(\tilde{x}_n^c, \lambda_o)} \tag{4.2}$$

converges to $x_L(\lambda)$ as $n \to \infty$. Here the correlation length is computed at the critical point defined using relation similar to ( 3.3).

In tables  4,5,6,7, we list the conformal weights computed at $\lambda_o = -3, -2, -1, 0$ for various $L$ and strip widths $n$. The extrapolated values for $n \to \infty$ are also given in the same tables. It is however not straight forward to compare these results with eqn.( 1.2) since the relation between $\lambda$ and $\lambda_o$ is not known except at the point where they both vanish. For this ordinary SAW point, the data are listed in table  7, here we clearly see that the extrapolated values agree very well with the exact values. In general convergence is better for sectors with small $L$.

In fig. 9, we plot successive values of $x_{1,n}(\lambda_o)$ against $\lambda_o$ for $L = 1$. Certain features are obvious from the figure: First, the conformal weight increases as $\lambda_o$ decreases as predicted by eqn.( 1.2). Note that $\lambda$ and $\lambda_o$ should have the same sign. Second, the crossings and minimums (for large $n$) of the lines seems to converge to the point $\lambda_o^*$ given in eqn.( 3.8). Furthermore, the value of $x_{1,n}(\lambda_o^*)$ at the crossing (minimum) converges to zero as $n$ increases. This supports the conjecture that the spiral transition takes place at $x_1(\lambda) = 0$[9]. From eqn.( 1.2), we deduce the renormalized $\lambda^*$ to be $5/96\pi$.

The vanishing of $x_1(\lambda_o^*)$ also implies that the exponent $\gamma$ at the spiral transition point is given by

$$\gamma = 2\nu_x \simeq 1.26 \pm 0.1 \tag{4.3}$$

using eqn.( 1.9). It is intriguing that such a relation between $\gamma$ and $\nu_x$ also exists for the ordinary SAW at the $\theta(\theta')$ point[13]. It should be stressed that $\gamma$ obtained in this way is certainly not rigorous, it would be desirable to check it against other methods. For example, the computation of the generating function

$$G(x, \lambda_o) = \sum_{|\vec{R}|} <\vec{S}(0) \cdot \vec{s}(\vec{R})> \sim |x - x^c|^\gamma \tag{4.4}$$

as $x$ approaches $x^c$.

One could also compute quantities that are independent of $\lambda$. For example, the following ratio

$$r_{L,L'}^{(n)}(\lambda_o) \equiv \left(\frac{n}{\pi \xi_{L,n}(\tilde{x}_n^c, \lambda_o)} + \frac{1}{12}\right) / \left(\frac{n}{\pi \xi_{L',n}(\tilde{x}_n^c, \lambda_o)} + \frac{1}{12}\right) \tag{4.5}$$

should converge to

$$\frac{x_L(\lambda) + 1/12}{x_{L'}(\lambda) + 1/12} = \left(\frac{L}{L'}\right)^2 \tag{4.6}$$

which depends only on the ratio of the number of 'legs'. However, because the size of our strip is small, successive $r_{L,L'}^{(n)}$'s do not form a monotonic sequence and extrapolation is difficult. We therefore compute the ratio $r_{L,L'}$ directly from the extrapolated values of the conformal weights given in tables 4,5,6,7. The results are presented in table 8 below. Convergence is better near $\lambda_o = 0$, nevertheless, we see evidence that the data agree to the exact values which are independent of $\lambda_o$.

## 5  Conclusion

In this numerical study, we verify several predictions and conjectures made in [9] regarding the repulsive regime. In particular, the critical exponent $x_L(\lambda)$ is checked and thermal exponent $\nu$ is found to have the value 3/4 independent of $\lambda$. Much less is known about the attractive regime except that $\nu$ is found to be 1/2 consistent with the fact that the polymer in this regime is dense.

The numerical data also give strong evidence that a spiral transition occurs as the current-current coupling $\lambda_o$ becomes large. The location of this transition is determined with less accuracy to be

$$\lambda_o^* \sim 1.1 \pm 0.2, \tag{5.1}$$

which is very close to the bound given in eqn.( 3.8). This could shed light on the nature of the spiral transition. On this transition point, we also deduce the exponents

$$\begin{aligned} \nu_x &= 0.63 \pm 0.05 \\ \gamma &= 1.26 \pm 0.10 \end{aligned} \tag{5.2}$$

which are different from that of the usual collapse transition of the ordinary SAW. This spiral transition certainly deserved further investigation.

Besides the critical line $x = x^c(\lambda)$, the region $x > x^c(\lambda)$ also deserves further investigation. For the ordinary SAW, it has been known for a long time that this region is critical and described by the low temperature phase of the $O(0)$ model[16], exact expressions for the critical exponents that have geometrical meaning have also been given. It would be interesting to study the effect of the current-current perturbation of this phase. We hope to pursue these questions in the future.

# Acknowlegement

I thank H. Saleur for suggesting this calculation and for the numerous illuminating conversations. Discussion with J.D. Noh on numerical analysis is gratefully acknowledged too. I also like thank the referee for pointing out the existence of a bound on $\lambda_o^*$ and numerous helpful comments.

# Figure Captions

- Figure 1: The two types of close encounter.

- Figure 2: Monomers and Connectivity.

- Figure 3: Susceptibility $c$ versus $\lambda_o$ for different strip widths.

- Figure 4: Density $\rho$ versus $\lambda_o$.

- Figure 5: $t$ versus $\lambda_o$ for different strip widths.

- Figure 6: Exponent $\nu$ against $\lambda_o$ for different strip widths $n$.

- Figure 7: Exponent $\nu$ against $\lambda_o$ for different strip widths $n$.

- Figure 8: Critical fugacity $x_n^c$ against $\lambda_o$ for different strip widths $n$.

- Figure 9: Conformal weight $x_{1,n}$ plotted against $\lambda_o$ for different strip widths $n$.

# Table Captions

- Table 1 Dimensions of various polymers sectors.

- Table 2 $\lambda_o^*$ obtained from maximums of successive $c_n$ and $t_n$ respectively.

- Table 3 $\lambda_o^*$ and $\nu_x$ obtained from crossings of lines in fig. 6 and fig. 7 respectively.

- Table 4 Conformal weights at $\lambda_o = -3$ from different strip widths and $L$.

- Table 5 Conformal weights at $\lambda_o = -2$ from different strip widths and $L$.

- Table 6 Conformal weights at $\lambda_o = -1$ from different strip widths and $L$.

- Table 7 Conformal weights at $\lambda_o = 0$ from different strip widths and $L$.

- Table 8 Ratio of conformal weights at various $\lambda_o$'s.

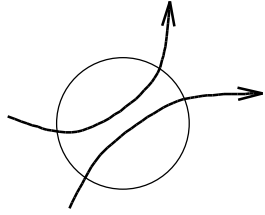
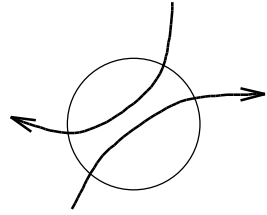

Parallel close encounter
weight: $e^{\lambda_\circ}$

Anti parallel close encounter
weight: 1

Figure 1: The two types of close encounter.

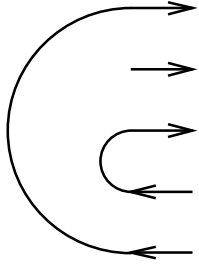

Figure 2: Monomers and Connectivity.

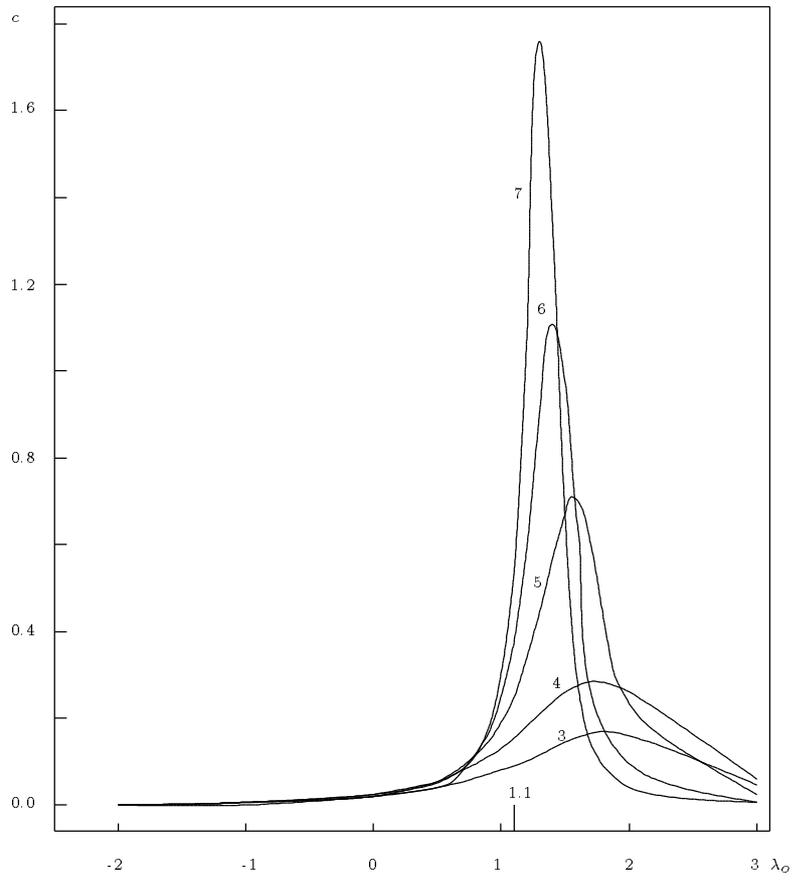

Figure 3: Susceptibility $c$ versus $\lambda_o$ for different strip widths.

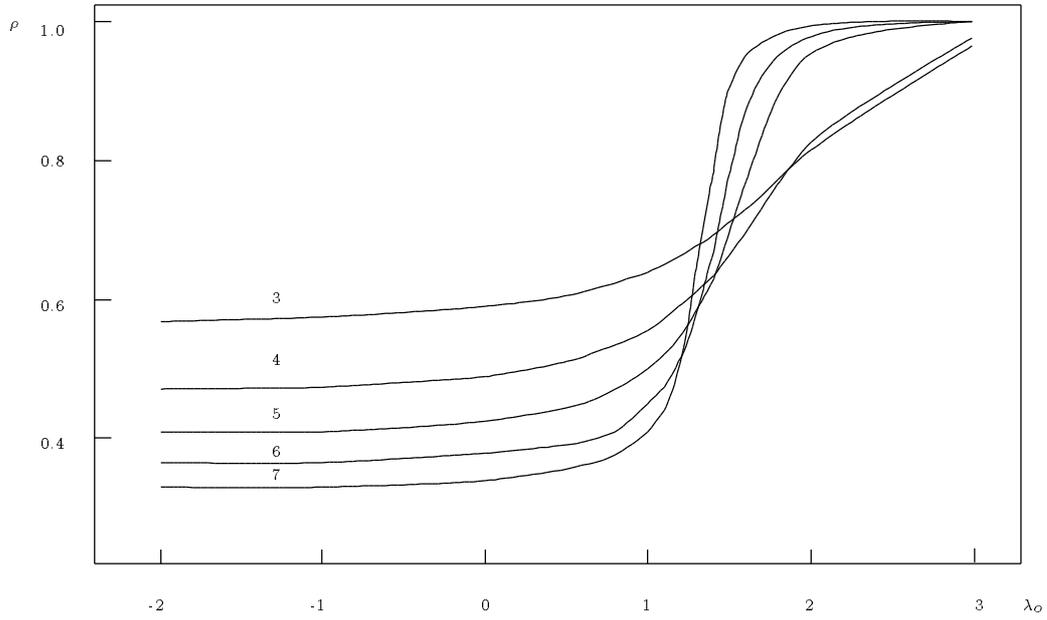

Figure 4: Density $\rho$ versus $\lambda_o$.

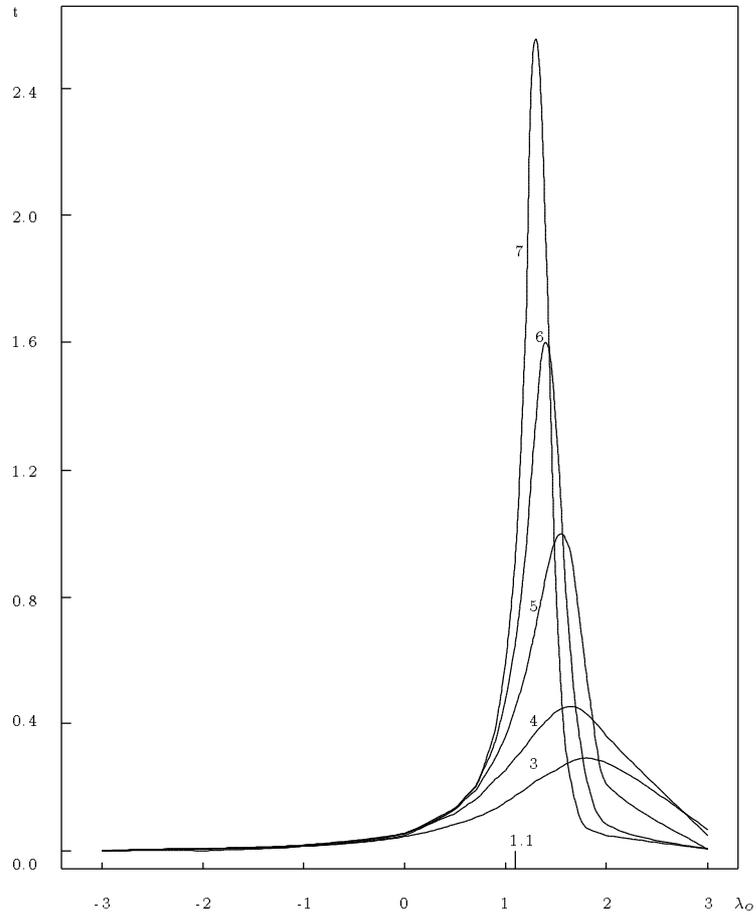

Figure 5: $t$ versus $\lambda_o$ for different strip widths.

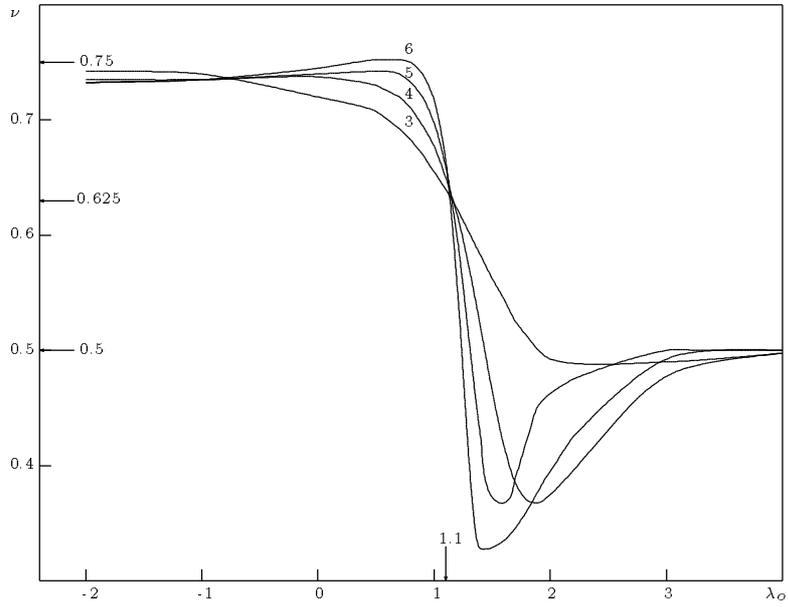

Figure 6: Exponent $\nu$ against $\lambda_o$ for different strip widths $n$.

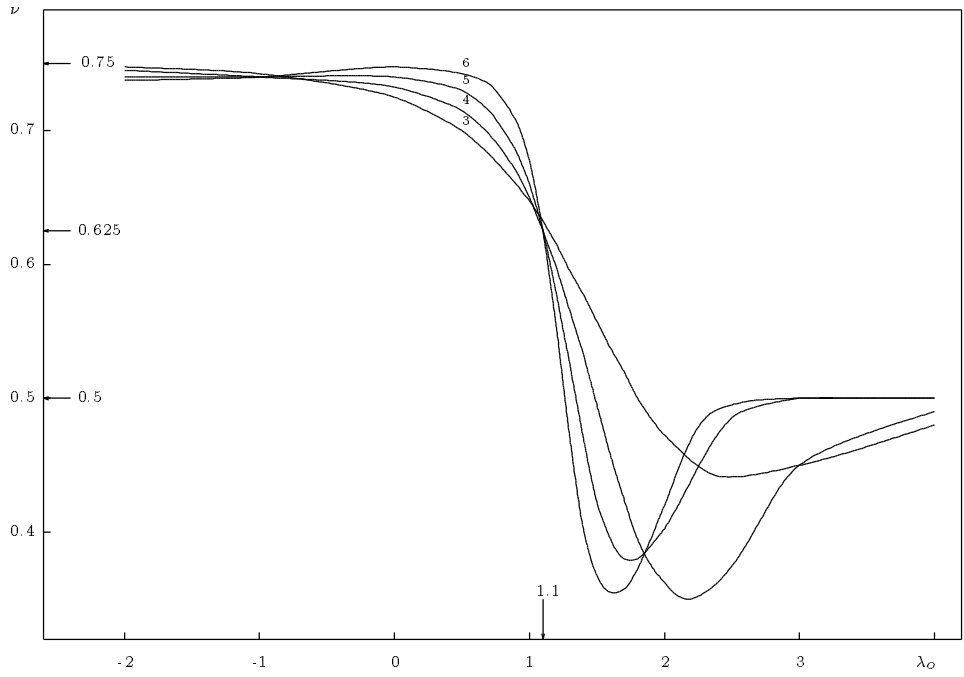

Figure 7: Exponent $\nu$ against $\lambda_o$ for different strip widths $n$.

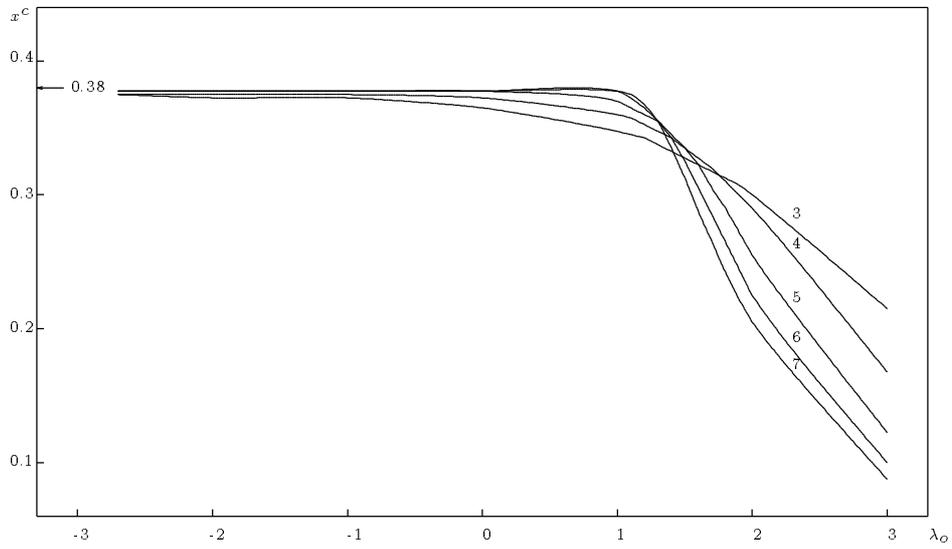

Figure 8: Critical fugacity $x_n^c$ against $\lambda_o$ for different strip widths $n$.

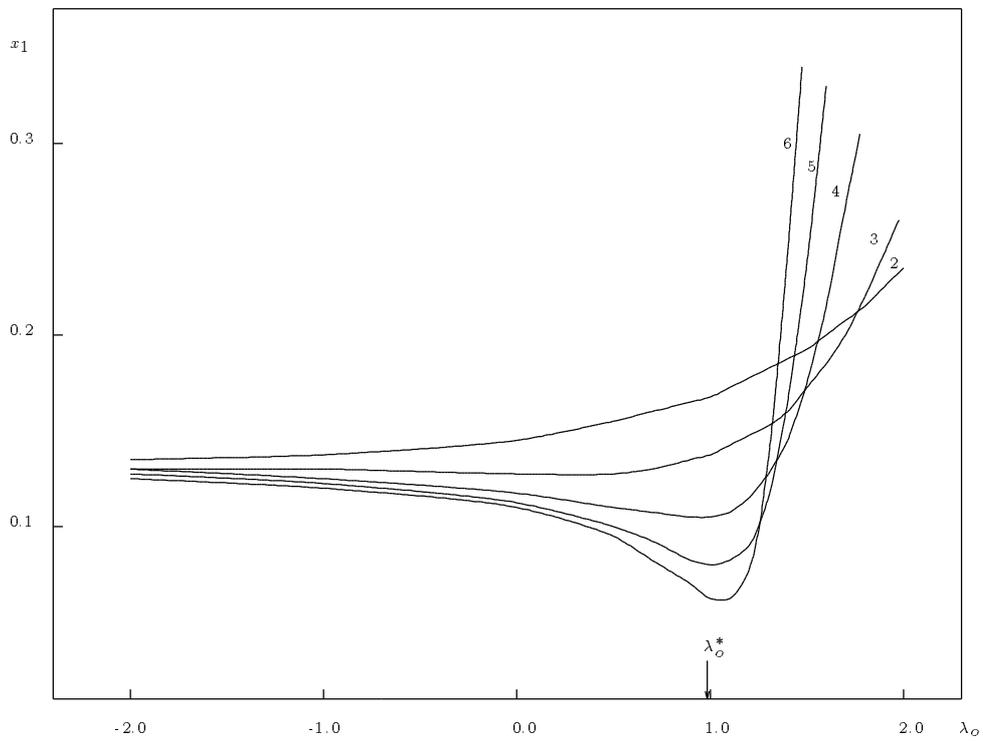

Figure 9: Conformal weight $x_{1,n}$ plotted against $\lambda_o$ for different strip widths $n$.

| $n$ | $S_{1,n}$ | $S_{2,n}$ | $S_{3,n}$ | $S_{4,n}$ | $S_{5,n}$ | $S_{6,n}$ |
|---|---|---|---|---|---|---|
| 2 | 3 | 1 | | | | |
| 3 | 11 | 3 | 1 | | | |
| 4 | 45 | 17 | 3 | 1 | | |
| 5 | 195 | 78 | 20 | 3 | 1 | |
| 6 | 881 | 402 | 122 | 26 | 3 | 1 |
| 7 | 4121 | 2017 | 689 | 171 | 29 | 3 |
| 8 | 19831 | 10470 | 3877 | 1114 | 231 | 35 |
| 9 | | | 21630 | 6802 | 1656 | 302 |
| 10 | | | | | 11140 | 2381 |

Table 1: Dimensions of various polymers sectors.

| $n$ | 3 | 4 | 5 | 6 | 7 | extrapolation |
|---|---|---|---|---|---|---|
| $\lambda_o^*$ | 1.85 | 1.75 | 1.57 | 1.41 | 1.32 | 1.17(20) |
| $\lambda_o^*$ | 1.82 | 1.65 | 1.53 | 1.41 | 1.33 | 1.12(20) |

Table 2: $\lambda_o^*$ obtained from maximums of successive $c_n$ and $t_n$ respectively.

| $n$ | 3,4 | 4,5 | 5,6 |
|---|---|---|---|
| $\lambda_o^*$ | 1.166 | 1.128 | 1.111 |
| $\nu_x$ | 0.628 | 0.642 | 0.641 |
| $\lambda_o^*$ | 1.016 | 1.084 | 1.087 |
| $\nu_x$ | 0.645 | 0.627 | 0.622 |

Table 3: $\lambda_o^*$ and $\nu_x$ obtained from crossings of lines in fig. 6 and fig. 7 respectively.

| $n$ | $x_{1,n}$ | $x_{2,n}$ | $x_{3,n}$ | $x_{4,n}$ |
|---|---|---|---|---|
| 2 | 0.12918 | 2.52743 | | |
| 3 | 0.12553 | 2.08935 | 5.68672 | |
| 4 | 0.12275 | 1.05229 | 5.31323 | 10.01097 |
| 5 | 0.12022 | 0.92181 | 4.12591 | 9.80083 |
| 6 | 0.11810 | 0.85923 | 2.61618 | 8.58174 |
| 7 | 0.11640 | 0.82220 | 2.32699 | 6.78043 |
| 8 | | | | 4.77810 |
| 9 | | | | 4.34188 |
| extrapolation | 0.116(8) | 0.736(10) | 1.731(9) | 3.91(11) |

Table 4: Conformal weights at $\lambda_o = -3$ from different strip widths and $L$.

| $n$ | $x_{1,n}$ | $x_{2,n}$ | $x_{3,n}$ | $x_{4,n}$ |
|---|---|---|---|---|
| 2 | 0.12918 | 1.89081 | | |
| 3 | 0.12479 | 1.61188 | 4.25433 | |
| 4 | 0.12186 | 1.01428 | 4.03999 | 7.56325 |
| 5 | 0.11936 | 0.90026 | 3.31607 | 7.41350 |
| 6 | 0.11732 | 0.84433 | 2.50402 | 6.66471 |
| 7 | 0.11570 | 0.81079 | 2.25849 | 5.61332 |
| 8 | | | | 4.57459 |
| 9 | | | | 4.20151 |
| extrapolation | 0.114(9) | 0.724(8) | 1.720(9) | 3.33(11) |

Table 5: Conformal weights at $\lambda_o = -2$ from different strip widths and $L$.

| $n$ | $x_{1,n}$ | $x_{2,n}$ | $x_{3,n}$ | $x_{4,n}$ |
|---|---|---|---|---|
| 2 | 0.12918 | 1.25419 | | |
| 3 | 0.12272 | 1.13442 | 2.82193 | |
| 4 | 0.11926 | 0.91246 | 2.78875 | 5.01677 |
| 5 | 0.11682 | 0.83990 | 2.47811 | 5.02618 |
| 6 | 0.11499 | 0.80148 | 2.20838 | 4.73237 |
| 7 | 0.11359 | 0.77747 | 2.06935 | 4.34551 |
| 8 | | | | 4.01605 |
| 9 | | | | 3.81281 |
| extrapolation | 0.112(9) | 0.703(8) | 1.690(8) | 3.22(9) |

Table 6: Conformal weights at $\lambda_o = -1$ from different strip widths and $L$.

| $n$ | $x_{1,n}$ | $x_{2,n}$ | $x_{3,n}$ | $x_{4,n}$ | $x_{5,n}$ |
|---|---|---|---|---|---|
| 2 | 0.12918 | 0.61757 | | | |
| 3 | 0.11651 | 0.65695 | 1.38954 | | |
| 4 | 0.11076 | 0.66641 | 1.49351 | 2.47029 | |
| 5 | 0.10809 | 0.66892 | 1.54045 | 2.63885 | 3.85982 |
| 6 | 0.10676 | 0.66948 | 1.56454 | 2.73158 | 4.09298 |
| 7 | 0.10601 | 0.66946 | 1.57813 | 2.78720 | 4.23650 |
| 8 | | | | 2.82302 | 4.33011 |
| 9 | | | | 2.84573 | 4.39399 |
| extrapolation | 0.1041(6) | 0.669(7) | 1.603(7) | 2.92(8) | 4.598(7) |
| exact | 0.10417 | 0.66667 | 1.60417 | 2.91667 | 4.60417 |

Table 7: Conformal weights at $\lambda_o = 0$ from different strip widths and $L$.

| $\lambda_o$ | $-3$ | $-2$ | $-1$ | $0$ | exact |
|---|---|---|---|---|---|
| $r_{2,1}$ | 4.11(14) | 4.08(13) | 4.02(16) | 4.01(12) | 4.00 |
| $r_{3,1}$ | 9.10(27) | 9.12(27) | 9.08(31) | 9.00(21) | 9.00 |
| $r_{4,1}$ | 17.44(89) | 17.28(84) | 16.90(83) | 16.02(66) | 16.00 |
| $r_{3,2}$ | 2.21(6) | 2.23(5) | 2.26(5) | 2.24(6) | 2.25 |
| $r_{4,2}$ | 4.24(15) | 4.22(13) | 4,20(12) | 3.99(12) | 4.00 |
| $r_{4,3}$ | 1.91(7) | 1.89(8) | 1.86(8) | 1.78(9) | 1.78 |

Table 8: Ratio of conformal weights at various $\lambda_o$'s.